\documentclass[aps,twocolumn,groupedaddress,elsart12,epsf]{revtex4-1}
\usepackage{amsmath}
\usepackage{epsfig,epsf}
\usepackage{graphicx}
\usepackage{verbatim}
\usepackage{epsfig,epsf}

\usepackage{pstricks}
\usepackage{color}

\begin{document}

\newcommand{\W} {\text{W}}
\def\beq{\begin{equation}}
\def\eeq{\end{equation}}
\newcommand{\nn}{\nonumber}

\def\beq{\begin{equation}}
\def\eeq{\end{equation}}
\newcommand{\bea}{\begin{eqnarray}}

\newcommand{\eea}{\end{eqnarray}}

\def\eq#1{{Eq.~(\ref{#1})}}
\def\fig#1{{Fig.~\ref{#1}}}


\title{Bose enhancement and the ridge}
\author{Tolga Altinoluk$^1$, N\'estor Armesto$^1$, Guillaume Beuf$^{2}$, Alex Kovner$^{3}$ and Michael Lublinsky$^2$}

\affiliation{
$^1$ Departamento de F\'{i}õsica de Part\'{i}culas and IGFAE, Universidade de Santiago de Compostela, 15706 Santiago de Compostela, Galicia-Spain\\
$^2$ Department of Physics, Ben-Gurion University of the Negev,
Beer Sheva 84105, Israel\\
$^3$ Physics Department, University of Connecticut, 2152 Hillside
Road, Storrs, CT 06269-3046, USA}

\begin{abstract}
We point out that Bose enhancement in a hadronic wave function generically leads to correlations between produced particles. We show explicitly, by calculating the projectile density matrix in the Color Glass Condensate approach to high-energy hadronic collisions, that  the Bose enhancement of gluons in the projectile leads to azimuthal collimation of long range rapidity correlations of the produced particles, the so-called ridge correlations.
\end{abstract}


\maketitle


\mbox{}

\pagestyle{plain}

\setcounter{page}{1}




\date{\today}

\section{Introduction}
The ridge structure observed in high multiplicity $p$-$p$ \cite{CMS} and $p$-Pb \cite{CMS:2012qk} collisions at the Large Hadron Collider (LHC)  triggered an intense activity aimed at understanding the possible physical origin of  correlations between emitted particles. Two basic ideas have been put forward in this context
(see others in \cite{Hwa:2008um}).

According to one idea, the origin of the correlations is the same as in similar ridge correlations observed earlier in  heavy ion collisions at the Relativistic Heavy Ion Collider \cite{PHOBOS} and the LHC \cite{Aamodt:2011by}. Namely, the angular collimation is due to flow effects in the final state \cite{Bozek:2012gr}. The qualitative features of the high multiplicity $p$-$p$ and $p$-Pb data, including the dependence on masses of produced particles, are well described by the hydrodynamic-based models. It is nevertheless
 challenging to explain how
the spatially small system produced in the final state in $p$-$p$ collisions can sustain the collective behavior necessary for local equilibration.

The second suggestion is that the final state correlations carry the imprint of the partonic correlations that  exist in the initial state. Three different variants of such initial state effects have been discussed in the literature: local anisotropy of target fields \cite{correlations}, spatial variation of partonic density \cite{LR} and finally the ``glasma graph'' contributions to particle production \cite{ddgjlv} within the Color Glass Condensate (CGC) approach to high-energy hadronic scattering. While the physical origin of the first two effects is quite clear, the physics behind glasma graphs has not been elucidated in the literature. On the other hand, numerical calculations based on the glasma graph approach have been very successful in reproducing the systematics of ridge correlations \cite{DV}.
It is therefore important to understand the  physics that underlies these numerical results.

The purpose of this Letter is to point out that there exists a general quantum mechanical mechanism that leads to positive correlations of emitted particles with similar quantum numbers. It is operative when the wave function of an incoming hadron is dominated by bosons (gluons), and is due to Bose enhancement in this wave function.
In the next section, after recalling the basic derivation, we will show that this is precisely the physical mechanism that underlies the glasma graph calculation of hadron production in $p$-$p$ and $p$-A collisions. The mechanism itself is however more general, has been widely used for identical mesons in heavy-ion collisions, see e.g. \cite{Weiner:1999th}.
Analogously, for fermions in the initial state one expects the opposite effect, namely Pauli blocking. In the final section, after discussing our results, we briefly address the question of which final state observables could be sensitive to the initial state Pauli blocking.

To avoid confusion, we stress that by Bose enhancement we do not mean the Hanbury-Brown-Twiss (HBT) correlations between emitted particles, which arise due to emission from a large source comprised of many incoherent emitters. We rather mean the effect of Bose statistics that enhances the probability to find two identical bosons with  the same transverse momentum {\it in the incoming projectile wave function before the collision}. Although this initial state enhancement is for two incoming bosons with identical transverse momentum, the momenta of the two are modified differently by the interaction with the target. Thus after scattering the two bosons emerge in the final state with different momenta. The correlation between the directions of the two momenta is nevertheless preserved in some range of kinematics, see later.

\section{Gluon production and Bose enhancement}
\subsection{Basics of Bose enhancement}

The prototypical textbook calculation of Bose enhancement proceeds as follows \cite{greiner}.
Consider a state with fixed occupation numbers of $N$ species of bosons at different momenta, $\vert \{n^i(p)\}\rangle\equiv\prod_{i,p}\frac{1}{\sqrt{n^i(p)!}} (a_i^{\dagger}(p)/\sqrt{V})^{  n^i(p)}\vert 0\rangle$, $i=1,\dots,N$, with a finite volume $V$ and periodic boundary conditions so that momenta are discrete.
The state is  translationally invariant with  mean particle density
\beq n\equiv \langle \{n^i(p)\}\vert a^{\dagger i}(x)a^i(x)\vert \{n^i(p)\}\rangle= \sum_{i,p}\,n^i(p).\eeq
Hereafter we take $\sum_p \approx \int d^3p/(2\pi)^3$. The 2-particle correlator in  coordinate space is
\beq D(x,y)\equiv\langle \{n(p)\}\vert a^{\dagger i}(x)a^{\dagger j}(y)a^{i}(x)a^j(y)\vert \{n(p)\}\rangle.\eeq
This is calculated by going to  momentum space, where the operator averages are simple:
\begin{eqnarray}\label{contr}&&\langle \{n(p)\}\vert a^{\dagger i}(p)a^{\dagger j}(q)a^i(l)a^j(m)\vert \{n(p)\}\rangle \\
&=&\langle \{n(p)\}\vert \delta(p-l)\delta(q-m)a^{\dagger i}(p)a^i(p)a^{\dagger j}(q)a^j(q) \nonumber \\&+&\delta(p-m)\delta(q-l)  a^{\dagger i}(p)a^j(p)a^{\dagger j}(q)a^i(q)\vert\{ n(p)\}\rangle\nonumber\\
&\approx& \delta(p-l)\delta(q-m) \sum_in^i(p)\sum_jn^j(q)\nonumber \\&+& \delta(p-m)\delta(q-l)\sum_in^i(p)n^i(q),\nonumber  \end{eqnarray}
where we have neglected the terms where all momenta are equal, which are suppressed by a phase space factor.
Using this, the result for $D(x,y)$ reads
\beq
D(x,y)=n^2+\sum_i\left\vert \int  \frac{d^3p}{(2\pi)^3}\; e^{ip(x-y)}n^i(p)\right\vert^2.\eeq
The last term expresses the Bose enhancement. It vanishes when the points are very far away, and gives ${\cal O}(1/N)$ enhancement when the points coincide. The ${\cal O}(1/N)$ suppression of the second term relative to the first one is  due to the fact that the second term contains a single sum over the species index. The physics is that only bosons of the same species are correlated with each other.
Technically the origin of this additional contribution
is the ``wrong contraction'' term in eq. (\ref{contr}).

The Bose enhancement is a generic phenomenon, and is not tied specifically to the state with fixed number of particles. An overwhelming majority of pure states or quantum density matrices exhibit Bose enhancement at some degree.
There is however one type of states that do not exhibit such behavior, notably classical-like coherent states. Consider a coherent state
\beq 
\vert b(x)\rangle\equiv \exp\{i\int d^3x\: b^i(x)(a^i(x)+a^{\dagger i}(x))\}\,|0\rangle.
\eeq
A trivial calculation in this state gives
\begin{eqnarray} &&\langle b(x)\vert a^{\dagger i}(x)a^i(x)\vert b(x)\rangle= b^i(x)b^i(x), \\ && \langle b(x)\vert a^{\dagger i}(x)a^{\dagger j}(y) a^i(x) a^j(y)\vert b(x)\rangle
=b^i(x)b^i(x)b^j(y)b^j(y), \nonumber\end{eqnarray}
so $D(x,y)=n(x)n(y)$.
Thus, in order to exhibit Bose enhancement, a state has to be nonclassical.

\subsection{Gluon production via glasma graphs}

As stated above, we want to demonstrate that the angular collimation arising from the glasma graph calculation owes its existence to the Bose enhancement in the projectile wave function.
Following \cite{DV,DMV}, we consider the calculation of inclusive two particle production
and assume local parton-hadron duality, namely that at a given momentum the number of produced hadrons  is proportional to the number of produced gluons.

\begin{figure*}[htp]
\includegraphics[width=\textwidth]{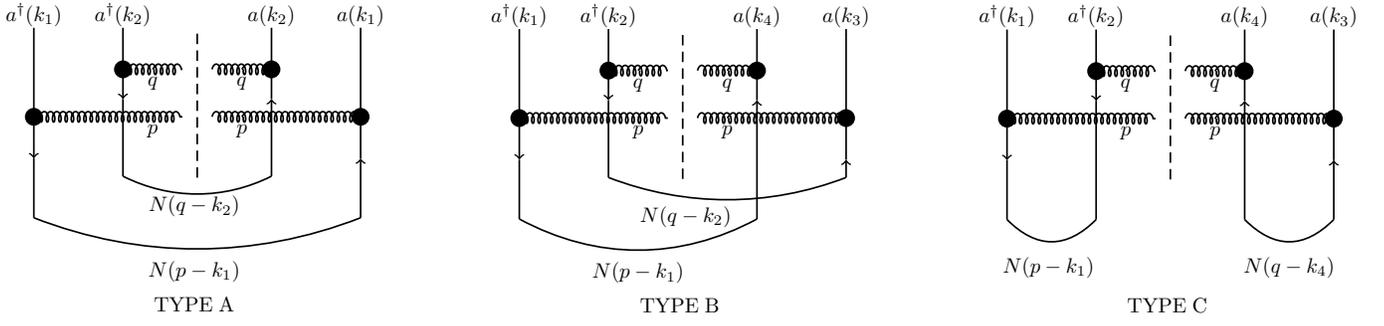}
\caption{\label{4g}Glasma graphs for two gluon inclusive production before averaging over the projectile color charge density $\rho$. Black blobs denote vertices which involve two momenta, e.g. $\delta^{ij}-{p^ik_{1}^{j}}/{p^2}$, and dashed lines the cuts. For details see the text.}
\end{figure*}

The graphs that contribute to this observable can be conveniently represented in terms of averages of gluon creation and annihilation operators in the incoming projectile wave function (see supplementary material and \cite{DMV}).
They come in three varieties, see fig. \ref{4g}. Type A graphs give the contribution whereby two gluons from the incoming projectile wave function scatter independently on the target. The incoming gluons have transverse momenta $k_1$and $k_2$ respectively. While propagating through the target the first particle picks up transverse momentum $p-k_1$ and the second particle picks up transverse momentum $q-k_2$, so that the outgoing particles have momenta $p$ and $q$. 
Type B and C graphs from the projectile point of view are ``interference graphs'', in the sense that the final state gluon with momentum $p$ comes from the projectile gluons with different momenta in the amplitude and complex conjugate amplitude.
Type B and C diagrams contain leading contributions that can be reinterpreted as Type A but with gluons originating from the target rather than from the projectile, and additionally subleading contributions, including those that lead to HBT correlations \cite{yury}. This is explained in detail in the Appendix. Therefore, in the following we will only discuss those of Type A, keeping in mind this complementary interpretation of the leading pieces of Type B and  C.

 
The Type A contribution to double inclusive gluon production can be written as
\begin{eqnarray}
&&C\int_{k_1,k_2}
\langle in\vert a^{\dagger i}_a(k_1)a^{\dagger j}_b(k_2)a^k_a(k_1)a^l_b(k_2) \vert in\rangle \\ &\times&\left[\delta^{ik}-\frac{k_{1}^{i}k_{1}^{k}}{p^2}\right]\left[\delta^{jl}-\frac{k_{2}^{j}k_{2}^{l}}{q^2}\right]N(p-k_1)N(q-k_2),\nonumber\label{twog}
\end{eqnarray}
where $|in\rangle$ is the wave function of the incoming projectile, $C$ is a constant,  $N(p-k)$ is the probability that the incoming gluon with transverse momentum $k$ acquires transverse momentum $p$ after scattering and, hereafter, we use the notation $\int_k\equiv \int \frac{d^2k}{(2\pi)^2}$. This scattering probability is of course determined by the distribution of target fields (within the glasma graph calculation, the scattering of the two gluons is independent). We have also assumed that the target wave function is translationally invariant, so that the momentum transfer is the same to the gluon in the amplitude and complex conjugate amplitude. The last assumption does not allow one to discuss the correlation mechanisms proposed in \cite{correlations,LR} within this framework.

Also note that we have not indicated the rapidity variable on the gluon creation and annihilation operators. Within the glasma graph calculation the gluon production is rapidity independent. Rapidity dependence becomes significant only when the rapidity difference between the observed hadrons becomes large, $\Delta \eta\sim 1/\alpha_s$.
The origin of this independence is that the CGC hadronic wave function is approximately boost-invariant. In fact, only the rapidity independent mode of the  gluon field is large in the wave function of the fast hadron, and only the creation operators of this one rapidity mode are relevant to the discussion of correlations.

Thus, the creation and annihilation operators entering the above equation are the original gluon operators integrated over rapidity,
\beq
a^i_a(k)\equiv\frac{1}{\sqrt Y}\int_{|\eta<Y/2|} \frac{d\eta}{2\pi} \  a^i_a(\eta,k).
\eeq
Here the rapidity interval $Y/2$ is arbitrary, but large enough to contain the rapidities of both  observed gluons. The operators defined this way satisfy the standard commutation relations in the transverse momentum space:
\beq[a^i_a(k),a^{\dagger j}_b(p)]=(2\pi)^2\delta_{ab}\delta^{ij} \delta^{(2)}(k-p).\eeq

The integral over momenta $k_1,k_2$ in eq. (\ref{twog}) contains a contribution from the region $k_1=k_2$. If the wave function $\vert in\rangle$ exhibits Bose enhancement, there is enhanced probability that the two gluons have equal momenta. This excess in the initial state will then translate into final state correlations. Note that this effect is suppressed by the squared number of colors $1/N^2_c$, since Bose enhancement is only operational for bosons with identical quantum numbers.

We thus have to understand what is the nature of the projectile state $\vert in\rangle$, and in particular we need to calculate
\beq D(k_1,k_2)\equiv\langle in\big\vert a^{\dagger i}_a(k_1)a^{\dagger j}_b(k_2)a^k_a(k_1)a^l_b(k_2) \big\vert in\rangle.
\eeq

Averaging over the projectile state in the standard CGC approach involves two elements. One needs to calculate the average over the soft degrees of freedom, as well as that over the valence color charge density. Conventionally this is done in the spirit of the Born-Oppenheimer approximation, namely first one averages over the soft gluon degrees of freedom at fixed valence color charge density $\rho$, and subsequently averages over the valence density distribution.

The wave function of the soft fields for fixed valence color charge density for a dilute projectile is a simple coherent state
\beq
\vert in\rangle_\rho=\exp\left\{i\int _k
 b^i_a(k)\left[a^{\dagger i}_a(k)+a^i_a(-k)\right]\right\}\vert 0\rangle,
\eeq
with the Weizs\"acker-Williams field $b^i_a(k)=g\rho_a(k)\frac{ik^i}{k^2}$.

 The averaging over the soft degrees of freedom leads to the well known expression for the observable in terms of the charge density:
\beq D(k_1,k_2)_\rho= b^i_a(k_1)b^j_b(k_2)b^k_a(-k_1)b^l_b(-k_2).\eeq
Since at fixed $\rho$, the soft gluon state is a coherent state, this expression does not seem to exhibit Bose enhancement. This is however misleading, since
 averaging over $\rho$ is part of the quantum averaging over the initial state wave function $\vert in\rangle$.
It is therefore instructive to reverse the conventional order of averaging, and average over the valence degrees of freedom first.
The result of such a procedure is a density matrix on the soft gluon Hilbert space. The subsequent averaging over this density matrix is a direct way to find out whether the projectile wave function exhibits  Bose enhancement.

\subsection{The soft gluon density matrix}
The soft gluon density matrix of course depends on the weight for the valence color charge density. For illustrative purposes we choose the same Gaussian weight used in the glasma graph calculation, the McLerran-Venugopalan model \cite{mv},
\beq
\langle \cdots\rangle_\rho={\cal N} \int D[\rho]\, \cdots \, e^{-\int_k
\frac{1}{2\mu^2(k)}\, \rho_a(k)\rho_a(-k)},
\eeq
where ${\cal N}$ is the normalization factor.

Thus the density matrix of the soft gluons is given by
\begin{eqnarray}
 \hat \rho&=&{\cal N}\int D[\rho] \;  e^{-\int_k
 \frac{1}{2\mu^2(k)}\rho_a(k)\rho_a(-k)} \\ &\times& e^{i\int_q
 b^i_b(q)\phi^i_b(-q)}\vert 0\rangle\langle 0\vert \; e^{-i\int_p
 b^j_c(p)\phi^j_c(-p)}\nonumber
\end{eqnarray}
where we have defined $
\phi^i_a(k)=a^i_a(k)+a^{\dagger i}_a(-k)$.
The integral over $\rho$ can be performed with the result
\begin{eqnarray}
\label{densitymatrix}
\hat\rho &=& e^{-\int_k
 \frac{g^2\mu^2(k)}{2k^4}k^i k^j\, \phi^i_b(k)\, \phi^j_b(-k)}  \\ &\times&\Bigg\{\sum_{n=0}^{+\infty}\frac{1}{n!} 
 \left[\prod_{m=1}^n \int_{p_m}
 \frac{g^2\mu^2(p_m)}{p_m^4}\: p_m^{i_m}\, \phi_{a_m}^{i_m}(p_m) \right]  \vert 0\rangle \nonumber\\
&\times &\langle 0\vert \left[\prod_{m=1}^n  p_m^{j_m}\phi_{a_m}^{j_m}(-p_m)   \right]\Bigg\}\nonumber \\
&\times& e^{-\int_{k^\prime}
\frac{g^2\mu^2(k')}{2{k'}^4}{k'}^{i'} {k'}^{j'}\, \phi_c^{i'}(k')\, \phi_c^{j'}(-k')}.\nonumber
\end{eqnarray}


The interesting correlator is given by
\begin{equation}  D(k_1,k_2)=tr[\hat \rho a^{\dagger i}_a(k_1)a^{\dagger j}_b(k_2)a^k_a(k_1)a^l_b(k_2)].
\end{equation}

It is a matter of straightforward algebra to show that
\begin{eqnarray}
&&tr [\hat \rho a^{\dagger i}_a(k)a^j_b(p)]=(2\pi)^2\delta_{ab}\; \delta^{(2)}(k-p)\; g^2\mu^2(p)\; \frac{p^ip^j}{p^4}\ ,\nonumber\\
&&tr [\hat \rho a^i_a(k)a^j_b(p)]=tr [\hat \rho a^{\dagger i}_a(k)a^{\dagger j}_b(p)]\nonumber \\ &=&-(2\pi)^2\delta_{ab}\;\delta^{(2)}(k+p)\; g^2\mu^2(p)\; \frac{p^ip^j}{p^4}
\end{eqnarray}
and then find
\begin{eqnarray}
&&D(k_1,k_2)=S^2(N_c^2-1)^2\frac{k_1^ik_1^kk_2^jk_2^l}{k_1^2k_2^2}\frac{g^4\mu^2(k_1)\mu^2(k_2)}{k_1^2k_2^2} \label{2part_result} \\
&&\times\left\{1+\frac{1}{S(N_c^2-1)}\left[\delta^{(2)}(k_1-k_2)+\delta^{(2)}(k_1+k_2)\right]\right\}.\nonumber 
\end{eqnarray}
In order to get eq. \eqref{2part_result}, we have made substitutions of the type $(2\pi)^2\delta^{(2)}(k_1-k_1)\to S$, where $S$ is the transverse area of the projectile. This regularization amounts to taking into account the discreteness of the transverse momentum spectrum of confined gluons.

\section{Discussion}
The first term in eq. (\ref{2part_result}) is the "classical" term equal to the square of the number of particles.
The second term is the typical Bose enhancement term, suppressed with respect to the first "`classical"' term by the total number of degrees of freedom (color and area).
The third term is specific to the density matrix at hand and, as explained in \cite{correlations}, appears due to reality of the gluon field scattering amplitude.
This establishes our point that the soft glue density matrix exhibits Bose enhancement, so that the likelihood of finding two gluons with the same transverse momentum is higher than average. Note that this effect is naturally subleading in $N_c$ as the enhancement is only effective if both gluons are in the same color state.

As a typical Bose enhancement contribution, the second term in eq. (\ref{2part_result}) is nonvanishing only when the momenta of the two gluons are equal. Note however  that  $k_1$ and $k_2$ are {\it not} the momenta of observed gluons, but rather the momenta of gluons in the wave function of the incoming projectile. The two gluons then scatter on the target and acquire momenta $p$ and $q$ with the probability $N(p-k_1)N(q-k_2)$, as indicated in eq.  (\ref{twog}).
Nevertheless it is clear that in favorable kinematics the initial state correlation of eq. (\ref{2part_result}) also appears as a correlation between the final state particles. Consider for example a situation where the incoming projectile wave function has an intrinsic saturation momentum $Q_s$, and the momenta $p$ and $q$ are chosen to be of the same order as $Q_s$ i.e. $p^2\sim q^2\sim Q_s^2$. In such a situation the production amplitude is dominated by the contribution from  $|k_1|,|k_2|\sim Q_s$, since the gluon density of the projectile is dominated by those gluons close to saturation momentum. The delta function in the Bose enhancement term in eq. (\ref{2part_result}) is then smeared when convoluted with the scattering probability $N(p-k_1)N(q-k_2)$, but positive angular correlations between the directions of $\vec p$ and $\vec q$ clearly survive. Thus the initial correlation is transmitted to the final state, provided fragmentation and final state effects in $p$-$p$ and $p$-A collisions are small. On the other hand, when $|p|,|q|\gg Q_s$, the initial correlation is smeared out by the large momentum transfer from the target, and the correlation in the final state should disappear. These qualitative features are of course borne out by the numerical calculations of \cite{DV}.

One interesting question naturally follows on from the above discussion. Fermions in the initial state wave function surely experience Pauli blocking. One therefore may expect negative correlation between final state hadrons that originate from quarks or antiquarks in the initial state. Such correlation should exhibit anticollimation rather than collimation, and therefore a valley rather than a ridge at $\Delta \phi=0$.
Whether such a valley extends to large relative rapidities between the observed particles is a question that should be explored. Quark-antiquark pairs are present in the hadronic wave function within the CGC approach at the next to leading order in $\alpha_s$ via splitting of gluons. Since the gluonic wave function is boost invariant, the same is true for the quark and antiquark distribution. However, the main question  here is whether the fluctuations around some "mean field" are not too large to mask the correlations in rapidity event by event.
Another way of saying it, is to recall that in our discussion of gluons only a single rapidity independent mode of the quantum gluon field was large in the CGC wave function. As a result any correlation extended over large intervals in rapidity. Whether a similar effect dominates the quark wave function has to be investigated. Work on these questions is ongoing \cite{inprogress}.

Perhaps an even more pressing question to understand is whether such valleys can be observed experimentally, given that the quark contribution is suppressed by $\alpha_s$ relative to that of gluons. Here we see two possible avenues. One point is that, as opposed to gluon contribution to correlations, the quark contribution is not symmetric under $\Delta\phi\rightarrow -\Delta\phi$. It thus can generate a nonvanishing $v_3$ coefficient within the CGC approach. Such mechanism will be quite different from the one explored in \cite{vn} based on the idea of local anisotropy suggested in \cite{correlations}.
Another possibility is to trigger on final states which predominantly arise from quarks. For example it may be interesting to study correlations between two D-mesons (or B-mesons), since open charm (or beauty) should have a relatively larger component coming from fragmentation of quarks, rather than that of gluons \cite{inprogress}.


\section*{Acknowledgments}
We express our gratitude to the Department of Physics of Ben-Gurion University of the Negev, for warm hospitality during stays when parts of this work were done (TA, NA and AK) and for financial support as Distinguished Scientist Visitor (NA). 
 ML thanks the Physics Department of the University of Connecticut for hospitality. 
This research  was supported by the People Programme (Marie Curie Actions) of the European Union's Seventh Framework Programme FP7/2007-2013/ under REA
grant agreement \#318921; the DOE grant DE-FG02-13ER41989 (AK); the BSF grant \#2012124 (ML and AK); the Kreitman Foundation (GB);  the  ISRAELI SCIENCE FOUNDATION grant \#87277111 (GB and ML);  the European Research Council grant HotLHC ERC-2011-StG-279579, Ministerio de Ciencia e Innovaci\'on of Spain under project FPA2014-58293-C2-1-P, Xunta de Galicia (Conseller\'{\i}a de Educaci\'on and Conseller\'\i a de Innovaci\'on e Industria - Programa Incite),  the Spanish Consolider-Ingenio 2010 Programme CPAN and  FEDER (TA and NA).

\section*{Appendix}
Here we explain in more detail the meaning of fig. \ref{4g}.
The diagrams   are depicted before averaging over the projectile wave function. It is however straightforward to identify the correspondence between the diagrams on fig. \ref{4g} and the glasma graphs of \cite{DMV}. The two possible nontrivial Wick contractions between $a$'s and $a^\dagger$'s in the diagram of Type A generate  the graphs $7a$ and $7b$ of \cite{DMV}. The contractions $\langle a^\dagger(k_1)a(k_3)\rangle \langle a^\dagger(k_2)a(k_4)\rangle$ in Type B  and C generate the glasma graphs (not explicitly drawn in \cite{DMV}) which  have the target and projectile charge densities interchanged. They can be interpreted as Type A but with gluons originating from the target rather than from the projectile. 
The contraction $\langle a^\dagger(k_1)a(k_4)\rangle\langle a^\dagger(k_2)a(k_3)\rangle$ in Type B and $\langle a^\dagger(k_1)a^\dagger(k_2)\rangle\langle a(k_3)a(k_4)\rangle$ in Type C diagrams generate diagram $8c$ and $8b$ of \cite{DMV} respectively, whose contribution is proportional to $\delta^{(2)}(p\mp q)$. As discussed in \cite{yury}, these diagrams are of the type that lead to HBT correlations if the translational invariance approximation is relaxed. Finally the contraction $\langle a^\dagger(k_1)a^\dagger(k_2)\rangle\langle a(k_3)a(k_4)\rangle$ in Type B and $\langle a^\dagger(k_1)a(k_4)\rangle\langle a^\dagger(k_2)a(k_3)\rangle$ in Type C give the diagram $8a$ of \cite{DMV}. The contribution of this diagram, as discussed in \cite{DMV}, is suppressed at high momenta. Although diagrams $8a,\ 8b$ and $8c$ were included in the numerical calculations of \cite{DV} for completeness, they  are 
subleading at high momentum and do not contribute significantly to ridge correlations. We will therefore disregard these contributions in the following.

 Recall that we are calculating
\begin{eqnarray}
&&\langle out\vert a^{\dagger i}_a(p)a^{\dagger j}_b(k)a^i_a(p)a^j_b(k)\vert out\rangle=\int D\rho W[\rho]\nonumber\\
&&\times\langle in\vert_\rho \hat S^\dagger e^{i\int _k
 b^i_a(k)\left[a^{\dagger i}_a(k)+a^i_a(-k)\right]}
a^{\dagger i}_a(p)a^{\dagger j}_b(q)a^i_a(p)a^j_b(q)\nonumber\\
&&\times e^{-i\int _k
 b^i_a(k)\left[a^{\dagger i}_a(k)+a^i_a(-k)\right]} \hat S\vert in\rangle_\rho\ .
 \end{eqnarray}
We now transform the creation and annihilation operators by the sequence of unitary transformations $e^{-i\int _k
 b^i_a(k)\left[a^{\dagger i}_a(k)+a^i_a(-k)\right]} \hat S$. The first transformation shifts the creation and annihilation operators by the Weizsacker-Williams field $b_a^i(k)$, while the second one color rotates the gluon operators as well as the color charge density operators locally in the transverse space. In coordinate space representation,
 \begin{eqnarray}
 &&\hat S^\dagger e^{i\int _k
 b^i_a(k)\left[a^{\dagger i}_a(k)+a^i_a(-k)\right]}
a^{\dagger i}_a(x)e^{-i\int _k
 b^i_a(k)\left[a^{\dagger i}_a(k)+a^i_a(-k)\right]} \hat S\nonumber\\
 &&=S_{ab}(x)a^{\dagger i}_b(x)-g\frac{\partial^i}{\partial^2}(x-y)S_{ab}(y)\rho_b(y)\,.\end{eqnarray}
We now use the fact that when acting on the state $\vert in\rangle_\rho$, the charge density operator and the gluon creation (and annihilation) operators are 
identical in the sense that $\left[\partial^ia^{\dagger i}_a(x)-\rho_a(x)\right]\vert in\rangle_\rho=0$.
We can thus write in momentum space
\begin{eqnarray}
&&\hat S^\dagger e^{i\int _k
 b^i_a(k)\left[a^{\dagger i}_a(k)+a^i_a(-k)\right]}
a^{\dagger i}_a(p)e^{-i\int _k
 b^i_a(k)\left[a^{\dagger i}_a(k)+a^i_a(-k)\right]} \hat S\nonumber\\
&&\times \vert in\rangle_\rho=
\int_{k_1}S_{ab}(p-k_1)\left[\delta^{ij}-\frac{p^ik_{1}^{j}}{p^2}\right]a^{\dagger j}_b(k_1)\vert in\rangle_\rho\ .\end{eqnarray}
Introducing the notation 
\begin{equation}\label{vert}
S_{ab}^{ij}(p,k)=S_{ab}(p-k)\left[\delta^{ij}-\frac{p^ik^j}{p^2}\right],
\end{equation}
we can now write the two gluon inclusive cross section in terms of the expectation value in the initial projectile state as
\begin{eqnarray}
 &&\int_{\{k_i\}}\langle S_{a a_1}^{i i_1}(p,k_1)S_{b a_2}^{ji_2}(q,k_2)S_{a a_3}^{ii_3}(-p,-k_3)S_{ba_4}^{ji_4}(-q,-k_4) \rangle \nonumber\\
&& \times \langle in\vert a_{a_1}^{\dagger i_1}(k_1)  a_{a_2}^{\dagger i_2}(k_2) a_{a_3}^{i_3}(k_3)a_{a_4}^{i_4}(k_4) \vert in\rangle.
\end{eqnarray}
The next step is to average over the target  fields. Within the glasma graph model one assumes that the averages of the $S$-matrices are translationally invariant, pairwise factorize, and the color structure of the averages is the same as in the dilute limit:
\begin{equation}
\langle S_{ab}(p)S_{cd}(k)\rangle=f^{abe}f^{cde}N(p)\delta(p+k).
\end{equation}
In this approximation the averaging over the target fields can be performed. This averaging leads to graphs of three distinct topologies depicted in fig. \ref{4g}. The blobs denote the appropriate vertices, as in eq. (\ref{vert}). For example, the two distinct blobs in the Type A diagram are $\delta^{ij} -\frac{p^ik_{1}^{j}}{p^2}$ and $\delta^{ij}-\frac{q^ik_{2}^{j}}{q^2}$.

\end{document}